# Self-avoiding random surfaces with fluctuating topology

M. Caselle[a], F. Gliozzi[a] and S. Vinti[a, b]

[a]Dip. di Fisica Teorica dell'Università di Torino and I.N.F.N., Via P. Giuria 1, I-10125 Torino, Italy

[b]Centro Brasileiro de Pesquisas Fisicas, Rua Dr. Xavier Sigaud 150, 22290 Rio de Janeiro, Brazil

A gas of self-avoiding surfaces with an arbitrary polynomial coupling to the gaussian curvature and an extrinsic curvature term can be realized in a three-dimensional Ising $bcc$ lattice with three local couplings. Similar three parameter realizations are valid also in other lattices. The relation between the crumpling transition and the roughening is discussed. It turns out that the mean area of these surfaces is proportional to its genus.

## 1. Introduction

It is known that a gas of self-avoiding random surfaces[1,2] (SARS) with unconstrained topology and weighted by the usual area term $e^{-\beta A}$ [3] belongs to the same universality class of the Ising model[4,5]. We discuss the behaviour of this gas when other kinds of couplings are introduced , like the extrinsic curvature term or a generic polynomial in the Riemannian curvature.

There are some three-dimensional lattices where the self-avoidance constraint is automatically satisfied and the SARS are simply the boundaries of the spin clusters[6,5,7,8]. In particular, in the body centered cubic lattice ($bcc$), it is possible to establish a one-to-one correspondence between the configurations of Ising spins and those of a gas of SARS in the dual lattice $\Lambda$ in the following way: for each pair of nearest neighboring sites $< i, j >$ with opposite spins $\sigma_i = \pm 1$ , $\sigma_i \neq \sigma_j$ we draw the dual plaquette, which is a hexagon, and similarly for each pair of next to the nearest neighboring sites $\ll i, j \gg$ of opposite spins we draw the dual plaquette which now is a square. The special property of the $\Lambda$ lattice is that each link belongs to only three plaquettes and this allows to enforce the self-avoidance constraint. As a consequence, each cluster of like spins is bounded by an oriented self-avoiding surface made with hexagons and squares. For instance, the surface associated to a cluster of a single site is a polyhedron of 14 faces which is nothing but the Wigner-Seitz cell of the $bcc$ Bravais lattice. Any connected aggregate of these polyhedra defines a self-avoiding surface. The whole $\Lambda$ lattice is obtained by filling the space with such polyhedra.

Each potential vertex of a $SARS$ of $\Lambda$ is dual to a tetrahedron of the $bcc$ lattice selected by a pair of orthogonal, adjacent links belonging to the two cubic sublattices (see fig.1).

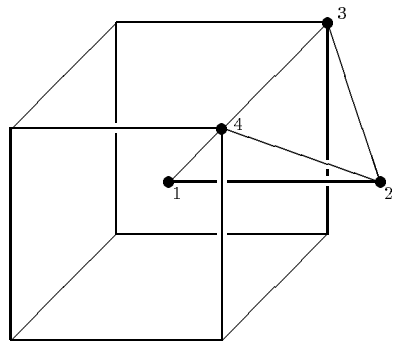

Figure 1. The tetrahedron 1234 of the $bcc$ lattice is dual to a potential vertex of a SARS on the dual lattice. The bonds $\ll 1, 2 \gg$ and $\ll 3, 4 \gg$ are dual to squares, while the bonds connecting nearest neighboring sites, like for instance $< 2, 3 >$ or $< 1, 4 >$, are dual to hexagons.

Since these tetrahedra fill the whole space, we can write the Hamiltonian of the Ising model as the sum of their contributions. On the other hand, the spin configurations of each tetrahedron can be used to read out the geometrical properties of the surface.

In particular, the Riemannian curvature of these polyhedric surfaces is obviously concentrated on the vertices. The contribution $R_v$ to the curvature of the vertex $v$ dual to the tetrahedron drawn in fig.1 can be written as a function of the spin configuration of the four sites 1,2,3,4:

$$24\,R_v = \sigma_{13} + \sigma_{14} + \sigma_{23} + \sigma_{24} - 3\sigma_{1234} - 1 \quad (1)$$

where $\sigma_{ij} = \sigma_i\sigma_j$. Note that when the four vertices 1,2,3,4 belong to the same cluster then $R_v = 0$ as it should be.

Denoting by $S$ a generic, not necessarily connected SARS, by $N_s(S)$ and $N_h(S)$ the number of squares and hexagons belonging to $S$, and by $\chi(S)$ its Euler characteristic, where

$$\chi(S) = 2(\#connected\ comp. - \#handles) \quad (2)$$

the partition function $Z(\beta_s, \beta_h, \beta_t)$ of a gas of SARS can be written as

$$Z = \sum_S e^{\left[-\beta_s N_s(S) - \beta'_h N_h(S) - \beta_t \chi(S)\right]} \quad (3)$$

where, besides the area terms $\beta_s$ and $\beta_h$ there is also a "topological fugacity" $e^{-\beta_t}$.

By using the Gauss-Bonnet theorem

$$\int d^2\xi \sqrt{g} R = \sum_v R_v = \chi(S) \quad (4)$$

and the previous expression for $R_v$, such a partition function can be rewritten as an Ising model on the $bcc$ lattice:

$$Z(\beta_s, \beta_h, \beta_t) = \sum_{\sigma = \pm 1} e^{-\beta H} \quad (5)$$

with the (reduced) Hamiltonian given by

$$-\beta H = \frac{\beta_h}{2} \sum_{<ij>} \sigma_{ij} + \frac{\beta_s}{2} \sum_{\ll kl \gg} \sigma_{kl} + \frac{\beta_t}{8} \sum_v \sigma_v \quad (6)$$

where $\sigma_v = \sigma_i\sigma_j\sigma_k\sigma_l$ is the product of the spins of the four vertices of the tetrahedron dual to the potential vertex $v$, $\beta_h = \beta'_h - \frac{\beta_t}{2}$ and an uninteresting additive constant has been omitted.

## 2. Riemannian Curvature

A crucial point is that any power of the curvature $R_v$ can be expressed again as a suitable combination of the three kinds of couplings $< ij >$, $\ll kl \gg$ and $(ijkl)$, as it follows directly from eq. (1). Thus any partition function of SARS, weighted by an action $\mathcal{A}$ polynomial in the curvature $R$

$$\mathcal{A} = \int d^2\xi \sqrt{g}[\mu + \beta_t R + \alpha R^2 + \ldots] \quad (7)$$

where $\mu$ is the cosmologic constant, can be discretized as an Ising model on the $bcc$ lattice described by a partition function $Z(\tilde{\beta}_s, \tilde{\beta}_h, \tilde{\beta}_t)$ of the same functional form of eq.(5), where $\tilde{\beta}_s, \tilde{\beta}_h$ and $\tilde{\beta}_t$ are suitable functions of the parameters $\mu, \beta_t, \alpha, \ldots$. For instance, a term proportional to $R^2$ has been introduced in some models [9] of dynamically triangulated surfaces (DTS) in order to study the transition from the crumpled to the collapsed phase, which is favoured for $\alpha \ll 0$. If we add a $R^2$ term to eq.(6) the new couplings are

$$\tilde{\beta}_h = \beta_h + \frac{\alpha}{6} \quad (8)$$

$$\tilde{\beta}_s = \beta_s - \frac{\alpha}{18} \quad (9)$$

$$\tilde{\beta}_t = \beta_t - \frac{5\alpha}{36} \quad . \quad (10)$$

The constraint of self-avoidance forbids collapsing of our surfaces. Note however that there are values of $\alpha < 0$ and $\beta_t$ such that $\tilde{\beta}_h = 0$ and $\tilde{\beta}_t = 0$ where the lattice decouples into a pair of simple cubic lattices. In these sublattices self-intersecting,collapsed surfaces are allowed. Note that the numerical coefficients of the above equations do not have any universal meaning, being of course specific of the $bcc$ lattice. It is however worth noting that also in the other lattices where the self-avoidance constraint is automatically satisfied, one can describe a gas SARS coupled to an arbitrary polynomial of the curvature $R$ in term of an Ising model with only three kinds of couplings. This is obvious in the lattice defined in [6], once one realizes that it is topologically dual to a $bcc$ lattice in a disguised form. It is also true in the dual lattice $\mathcal{L}$ of the face centered cubic ($fcc$) lattice, used in ref.[5], where apparently there are more complicated couplings. Indeed, there are two kinds of potential vertices for SARS of $\mathcal{L}$ : one is specified by the spins of four suitable neighboring sites, while the other needs the spin status of eight sites. Denoting respectively by $R_v^t$ and $R_v^o$ the contribution to the curvature of these vertices and using formulas similar to eq.(1), one easily verifies that

$$(R_v^t)^2 = \frac{1}{4} R_v^t \tag{11}$$

and

$$(R_v^o)^2 = -\frac{1}{2} R_v^o \quad . \tag{12}$$

Hence, a gas of SARS described by the action (7) may also be discretized as a Ising model in the $fcc$ lattice with three kinds of couplings associated to $R_v^t$, $R_v^o$ and to the pairs of nearest neighboring sites.

## 3. Extrinsic Curvature

Coming back to the $bcc$ lattice, we want to show that it is also possible to put in the same three-parameter form a SARS partition function with an extrinsic curvature term

$$E = \sum_{(a,b)} (1 - \cos(\vartheta_{a,b})) \tag{13}$$

where $\vartheta_{a,b}$ denotes the angle between two incident faces $a$ and $b$ . In the $\Lambda$ lattice there are only two kinds of such angles: if $a$ and $b$ are both hexagons one finds $\cos(\vartheta_{a,b}) = -\frac{1}{3}$ ; if one of them is a square, then $\cos(\vartheta_{a,b}) = \frac{1}{\sqrt{3}}$.

Modifying the partition function $Z(\beta_s, \beta_h, \beta_t)$ of eq.(6) as follows

$$Z_E(\tilde{\beta}_s, \tilde{\beta}_h, \tilde{\beta}_t) = \sum_{\sigma = \pm 1} e^{-\beta H - \lambda E} \tag{14}$$

where $e^{-\lambda}$ is the fugacity associated to the extrinsic curvature term, we find

$$\tilde{\beta}_h = \beta_h + \frac{\lambda}{3} \tag{15}$$

$$\tilde{\beta}_s = \beta_s + c\lambda \quad , \quad c = \frac{32\sqrt{3} - 21}{24} \tag{16}$$

$$\tilde{\beta}_t = \beta_t - \frac{11\lambda}{6} \quad . \tag{17}$$

In the DTS models the coupling to $E$ is used to control the crumpling transition [10] : $\lambda$ drives the system from a crumpled phase to a smooth phase. We see from the above expressions that the flow towards the smooth surfaces implies a cooling of the gas. It is conceivable that the crumpling transition of the single surface would correspond to the roughening point of the Ising model. A numerical analysis in this direction might be very interesting.

## 4. Condensation of Handles

Microscopic analysis of the topology of the SARS near the critical point of the Ising model by directly counting the number of handles of the connected boundaries of the spin clusters has been performed only recently [8,11].

It has been found for different lattices and in a wide range of temperatures near the critical point of the Ising model, that the number $\mathcal{N}$ of SARS as a function of the area $A$ and the genus $G$ is accurately described for large surfaces by [8,11]

$$\mathcal{N} \propto A^{f(G)} e^{-\mu A} \tag{18}$$

where the "cosmological constant" $\mu$ is a function of the lattice shape and $f(G)$ is an almost linear function of the genus. It has been pointed out [11] that SARS with unconstrained topology can also be reconstructed from the spin configurations of a simple cubic lattice, even if it is well known that in such a lattice the cluster boundaries do not satisfy the self-avoiding constraint, because of the possible presence of lines of self-contact. These lines are the only potential sources of ambiguities in the topological reconstruction of the surface. By small deformations of these surfaces it is possible to eliminate, at least locally, these contact lines. However, if these deformations are chosen arbitrarily it may appear global obstructions in the surface reconstruction. It has been found a set of rules to disentangle in a simple way these contact lines which avoids global obstructions and assigns to each Ising configuration a set of SARS [11]. Notice that this set of rules is by no means unique: there are other sets of consistent rules yielding for the same Ising configuration a different set of SARS. Notwithstanding, the only modification observed in eq.(18) is a variation of the parameter $\mu$.

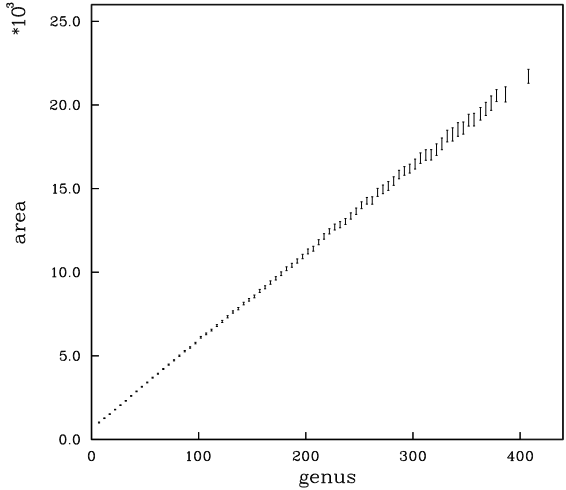

Figure 2. *The mean area of an interface as a function of its genus in a cubic lattice. See ref. [11].*

An important consequence of the linearity of $f(G)$ is the phenomenon of condensation of handles: the mean area $< A >_G$ of the surface at fixed genus is proportional to its genus. Indeed, by eq.(18)

$$< A >_G = -\frac{\partial}{\partial \mu} \log \mathcal{N} \qquad (19)$$

one gets at once

$$< A >_G = \frac{f(G)}{\mu} \approx \frac{b}{\mu} G + \frac{c}{\mu} \quad . \qquad (20)$$

Thus the number of handles per unit area is a non-vanishing constant $b/\mu$; this means that the SARS are dominated by configurations with a large number of microscopic handles, then the cluster boundary is much more similar to a sponge rather than to a smooth surface. See fig.2.

The universal parameter $b$ is related to the string susceptibility $\gamma_s$ by $b = 2 - \gamma_s$. The fact that all SARS in the *bcc* lattice can be generated by gluing together in all the possible ways the Wigner-Seitz cells suggests that they belong to the same universality class of the branched polymers, which gives $\gamma_s = \frac{1}{2}$. Preliminary numerical simulations suggest a slightly larger value.

There are arguments suggesting that SARS with unrestricted topology could be in a different universality class [12,13]; however recent numerical simulations on self-avoiding, flexible, dynamically triangulated surfaces seem to indicate that the number of handles is irrelevant when classifying SARS in universality classes [14]. It would be very interesting to evaluate the radius of gyration of SARS in the dual of *bcc* lattice in order to settle this point.

## REFERENCES

1. T.Sterling and J.Greensite, Phys. Lett. B 121 (1983) 345.
2. B.Durhuus, J.Fröhlich and T. Jonsson, Nucl. Phys. B 225 (1983) 183
3. M.Karowski and H.J. Thun, Phys. Rev. Lett. 54 (1985) 2556
4. M.Karowski, J. Phys. A 19 (1986) 3375
5. F.David, Europhys. Lett. 9 (1989) 581
6. T.A. Larsson, J.Phys. A 20 (1987) L-535
7. D.A. Huse, Phys. Rev. Lett. 64 (1990) 3200
8. V.S.Dotsenko, G. Harris, E. Marinari, E. Martinec, M. Picco and P. Windey, preprint EFI 93-24 (1993)
9. J. Ambjørn, B. Durhuus and J. Frölich, Nucl. Phys. B 275 (1986) 161
10. J.Ambjørn, B.Durhuus, J. Frölich and T. Jonsson, Nucl. Phys. B 290 [FS20] (1987) 253
11. M. Caselle, F. Gliozzi and S. Vinti, preprint DFTT 12/93 (1993)
12. J.R. Banavar, A. Maritan and A. Stella, Science, 252 (1991) 825
13. A. Maritan, F. Seno and A. Stella, Phys. Rev B 44 (1991) 2834
14. C. Jeppesen and J. H. Ipsen, Europhys. Lett. 22 (1993) 713